# Beam Shaping Based on Axisymmetric Aspheric Mirrors

**Zhihao Chen, Xiaonan Ning, Jiucheng Chen, Jianfei Hua* and Wei Lu**

*Department of Engineering Physics, Tsinghua University, Beijing 100084, China.*
*\*Corresponding author: jfhua@tsinghua.edu.cn*

Flat-top beam, known for its ability to generate a consistently even irradiation area, holds vast utility in many fields of scientific and industrial applications. In this paper, a reflective laser beam shaping method based on two axisymmetric aspheric mirrors (AAMs), a polarizing beam splitter (PBS) and two quarter wave plates (QWPs) is proposed to transform Gaussian beam into flat-top beam. Compared to alternative beam shaping methods, the method using AAMs demonstrates distinct advantages on notably high energy efficiency and unique capability to generate parallel beams. Thanks to its relative simplicities of design, manufacture and tunability, AAMs-shaping further enhances its appeal in applied research scenarios.
**Keywords**: beam shaping; aspheric mirror; Gaussian beam; flat-top beam
DOI:

## 1. Introduction

There is a significant demand for laser beams with flat-top intensity distributions in various fields, such as laser processing, nonlinear optics, high-energy laser amplification, and so on. These applications require a precisely shaped laser beam with a consistent power distribution, which make the flat-top beam being a highly desired tool for these tasks [1-4]. During laser cutting, the flat-top beam allows precise material removal in dissolution, resulting in sharp edges. As for the laser ablation, flat-top laser beam can evenly distribute the welding temperature and produce high quality welds. In nonlinear optical applications, such as in optical parametric chirped pulse amplification (OPCPA) systems, the use of a pump beam with a uniform distribution of light intensity can significantly enhance the efficiency of pump-to-signal conversion [5]. However, most laser resonators are operated on stable TEM00 mode, resulting in the generated laser beam with a spatial Gaussian distribution. Therefore, it is crucial to investigate how to obtain flat-top beam from Gaussian beam.

A great number of beam-shaping methods have been proposed to achieve flat-top laser beam like diffractive optical element (DOE) [6-9], spatial light modulator (SLM) [10-12], spatially variable wave plate (SVWP) [13-15], birefringent lenses [16], double free-form mirrors [17-18] (somewhat analogous to double aspheric lenses [19-20]) and microlens array, et al. The shaping methods of DOE and SLM are characterized by energy efficiency up to 80% [9,12], but cannot obtain collimated flat-top beam after reshaping. Alternatively, the methods of SVWP and birefringent lenses can maintain the phase of the laser beam, yet their energy efficiency is less than 60% [13,16]. Equipped the capability to produce a collimated beam, beam shaping with double free-form mirrors excel among all techniques due to ~100% energy efficiency without considering transmission loss or reflection loss.

For the laser beam shaping with a couple of free-form mirrors, the first mirror is employed to reshape the intensity distribution, while the second one acts as a phase corrector to collimate the beam [17]. When reshaping the laser beam using double free-form mirrors, the laser beam should deviate from its initial optical axis by introducing off-axis angles and the surfaces of both mirrors need to be non-axisymmetric. Consequently, the free-form mirrors are characterized without a rotational symmetry axis, resulting in the absence of specific geometric characteristics and predefined machining references. Compared to the processing of axisymmetric mirrors, the machining of free-form mirrors demands higher requirements on the machine tools. A five-axis ultra-precision lathe equipped with fast tool servo (FTS) technology is typically required to achieve the desired complexity and precision for fabrication of non-axisymmetric free-form mirrors [21]. In contrast, the fabrication of axisymmetric aspherical mirrors (AAMs) can be accomplished using a conventional two-axis ultra-precision lathe, employing grinding and polishing techniques.

In this paper, a beam shaping technique using AAMs is proposed and demonstrated to efficiently transform Gaussian beam into flat-top beam. For AAMs, the same beam shaping effect can be achieved based on a normal incidence to axisymmetric mirrors, when combining polarization manipulation and perpendicularly reflection of the laser beam. During laser alignment, the laser beam can be more easily controlled when incident perpendicular to the mirror, as compared to incident it at a specific angle for the case of double free-form mirrors.

This paper is structured as follows. In Sect. 2, the universal equation for the surface profiles of two free-form mirrors is derived for beam shaping. In Sect. 3, the reshaping from Gaussian beam to flat-top beam is further realized based on the curve equations of the two free-form mirrors in polar coordinate system. The above curve equations are simplified for AAMs in Sect. 4 and the numerical demonstration is carried out in Sect. 5. In Sect. 6, two processed AAMs are experimentally demonstrated to transform Gaussian beam into flat-top beam and a brief discussion is given in Sect.7.

## 2. Derivation of two surface equations

Since axisymmetric aspheric mirrors are specific cases of non-axisymmetric free-form mirrors, the generalized description of surface can be solved using two second-order partial differential equations (PDEs) of Monge–Ampère type [18]. If both the input and the output laser beam have axisymmetric intensity profiles, such as Gaussian beam and flat-top beam, the two PDEs above can be simplified to one ordinary differential equation (ODE).

Figure 1 illustrates a beam shaping system comprising of two free-form mirrors, designed specifically to transform Gaussian beam into flat-top beam. The two Cartesian coordinate system $(x, y, z)$ and $(X, Y, Z)$ are established respectively on two free-form mirrors, where the contact points serve as the origin and the incident direction of the laser beam are denoted as the z-axis and Z-axis, respectively. We assume that the origin of the Cartesian coordinate system $(x, y, z)$ in $(X, Y, Z)$ is $(a, b, c)$. Consequently, the transformation relationship between the two coordinates is $X = x - a$, $Y = y - b$ and $Z = z - c$.

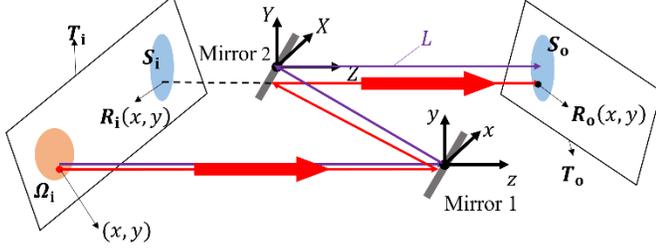

Fig. 1. Layout of two non-axisymmetric free-form mirrors. The purple line(L) represents the central ray of the input beam, while the red line represents any other ray of the input beam except for the central ray.

The input laser beam is located in the incident plane $T_i$, and the output laser beam is located in the output plane $T_o$. The input beam projected onto the plane $T_i$ is referred to as $\Omega_i$. Similarly, the projection of the output beam onto the plane $T_i$ is labeled as $S_i$, and its projection onto the plane $T_o$ is designated as $S_o$.

The point of intersection between the input beam and the output plane $T_i$ is denoted as $(x, y)$, while its intersection with $T_o$ is represented as $\mathbf{R_o}(x, y)$. Furthermore, the projection of $\mathbf{R_o}(x, y)$ onto $T_i$ is labeled as $\mathbf{R_i}(x, y)$. The equations that govern the curvature of the two free-form mirrors are solved subsequently to obtain the expressions for $z(x, y)$ and $Z(X, Y)$. The relationship among the coordinate projections $(x, y)$, $\mathbf{R_i}(x, y)$ and the curve equation $z(x, y)$ of the first mirror [18] can be established as

$$\mathbf{R_i}(x, y) = (x, y) + (\sqrt{a^2 + b^2 + c^2} + c) \cdot \nabla z \quad x \in \Omega_i \quad (1)$$

This PDE can be rewritten as

$$(X + a, Y + b) = (x, y) + (\sqrt{a^2 + b^2 + c^2} + c) \cdot \nabla z \quad x \in \Omega_i \quad (2)$$

Currently, we only require expressions for $X(x, y)$ and $Y(x, y)$. The equation for the curve of the first mirror can be got by integrating over $\nabla z$. As for the curve equation of the second free-form mirror, the coordinates $(x, y)$ and $\mathbf{R_i}(x, y)$ should have the same curvature at the corresponding points on both mirrors to guarantee the input and output laser beams are parallel. Therefore, the relationship between $\nabla z$ and $\nabla Z$ can be derived as follows

$$\begin{cases} \dfrac{\partial Z[X(x,y), Y(x,y)]}{\partial X(x,y)}\bigg|_{(x_0,y_0)} = \dfrac{\partial z(x,y)}{\partial x}\bigg|_{(x_0,y_0)} \\ \dfrac{\partial Z[X(x,y), Y(x,y)]}{\partial Y(x,y)}\bigg|_{(x_0,y_0)} = \dfrac{\partial z(x,y)}{\partial y}\bigg|_{(x_0,y_0)} \end{cases} \quad (3)$$

## 3. Reshaping Gaussian beam into flat-top beam

In the previous section, the equations $z(x, y)$ and $Z(X, Y)$ are acquired from $X(x, y)$ and $Y(x, y)$ for the two corresponding aspheric mirrors. In this section, the focus is on deducing the expressions for $X(x, y)$ and $Y(x, y)$ that reshape Gaussian beam into flat-top beam. By doing so, the expressions for $z(x, y)$ and $Z(X, Y)$ are obtained, which describe the curvature of the mirrors required to achieve this beam transformation.

The intensity distribution of Gaussian beam is

$$I_{in}(r, \theta) = I_1 \cdot \exp(-\frac{2r^2}{w^2}) \quad (4)$$

There are numerous functional forms that can be utilized to represent the intensity distribution of a flat-top beam, including the Super-Gaussian function, Flattened-Gaussian function, and Flattened-Lorentz function [22,23]. To facilitate function integration and variable separation, the Flattened-Lorentz function is selected as the fitting function for the intensity distribution of the flat-top beam. In Eq. 4 and 5, $I_1$ and $I_2$ denote the intensity of light beams at the interface centers of the input and output beams, respectively.

$$I_{out}(R, \Theta) = I_2 \cdot \frac{1}{\left[1 + \left(\dfrac{R}{R_{FL}}\right)^q\right]^{1+\frac{2}{q}}} \quad (5)$$

Here, $q$ is associated with the steepness of the edge of the Flattened-Lorentz function. As $q$ increases, the edge of the Flattened-Lorentz function becomes steeper. $R_{FL}$ corresponds to the radius of the flat-top beam. To meet energy conservation requirement, the integral of the laser beam intensity over the entire plane is normalized as

$$\int_0^{2\pi} d\theta \int_0^{+\infty} I_1 \cdot r \cdot \exp(-\frac{2r^2}{w^2}) dr = \int_0^{2\pi} d\Theta \int_0^{+\infty} I_2 \cdot \frac{R}{\left[1 + \left(\dfrac{R}{R_{FL}}\right)^q\right]^{1+\frac{2}{q}}} dR = 1 \quad (6)$$

The solutions to these two equations are $I_1 = 2/(\pi \cdot w^2)$ and $I_2 = 1/(\pi \cdot R_{FL}^2)$. The energy conservation for a fixed azimuth angle is employed to determine the coordinate correspondence between the two coordinate systems

$$\int_0^{\theta'} d\theta' \int_0^{r'} \frac{2r'}{\pi \cdot w^2} \cdot \exp(-\frac{2r'^2}{w^2}) dr' =$$
$$\int_0^{\Theta'} d\Theta' \int_0^{R'} \frac{1}{\pi \cdot R_{FL}^2} \cdot \frac{R'}{\left[1+\left(\frac{R'}{R_{FL}}\right)^q\right]^{1+\frac{2}{q}}} dR' \quad (7)$$

Thus, the relations between the input and output beams can be obtained from Eq. 7

$$\begin{cases} R = R_{FL} \cdot \left\{ -1 + \left[1 - \exp(-2 \cdot \frac{r^2}{w^2})\right]^{-\frac{q}{2}} \right\}^{-\frac{1}{q}} \\ \Theta = \theta \end{cases} \quad (8)$$

From Eq. 2, we can obtain

$$\begin{cases} \frac{\partial z(x,y)}{\partial x} = \frac{X(x,y) + a - x}{\sqrt{a^2 + b^2 + c^2} + c} \\ \frac{\partial z(x,y)}{\partial y} = \frac{Y(x,y) + b - y}{\sqrt{a^2 + b^2 + c^2} + c} \end{cases} \quad (9)$$

The coordinate conversion relation between Cartesian coordinate system and polar coordinate system is $x = r\cos\theta$, $y = r\sin\theta$, $X = R\cos\Theta$, $Y = R\sin\Theta$. The relationship between $(R, \Theta)$ and $(r, \theta)$ can be found in Eq. 8, therefor the differential equations in polar form are

$$\begin{cases} \frac{\partial z(r,\theta)}{\partial r} = \frac{R(r,\theta) - r + a\cos\theta + b\sin\theta}{\sqrt{a^2 + b^2 + c^2} + c} \\ \frac{\partial z(r,\theta)}{\partial \theta} = \frac{-ar\sin\theta + br\cos\theta}{\sqrt{a^2 + b^2 + c^2} + c} \end{cases} \quad (10)$$

The expression of $z(r,\theta)$ can be obtained by integrating the following equation as

$$z(r_0, \theta_0) = \int_{(0,0)}^{(r_0, \theta_0)} \frac{\partial z(r,\theta)}{\partial r} dr + \frac{\partial z(r,\theta)}{\partial \theta} d\theta$$
$$= \int_{(0,0)}^{(r_0, 0)} \frac{\partial z(r, 0)}{\partial r} dr + \int_{(r_0, 0)}^{(r_0, \theta_0)} \frac{\partial z(r_0, \theta)}{\partial \theta} d\theta \quad (11)$$

The distribution of the input beam can be obtained by inverting Eq. 8 as

$$\begin{cases} r = \sqrt{\frac{\ln\left\{1 - \left[\left(\frac{R}{R_{FL}}\right)^{-q} + 1\right]^{-\frac{2}{q}}\right\} \cdot w^2}{-2}} \\ \theta = \Theta \end{cases} \quad (12)$$

Likewise, the expressions of $\partial Z/\partial R$ and $\partial Z/\partial \theta$ are

$$\begin{cases} \frac{\partial Z(R,\Theta)}{\partial R} = \frac{R - r(R,\Theta) + a\cos\Theta + b\sin\Theta}{\sqrt{a^2 + b^2 + c^2} + c} \\ \frac{\partial Z(R,\Theta)}{\partial \Theta} = \frac{-aR\sin\Theta + bR\cos\Theta}{\sqrt{a^2 + b^2 + c^2} + c} \end{cases} \quad (13)$$

The expression of $Z(R,\Theta)$ can be obtained like Eq. 11.

4. Design of axisymmetric free-form mirrors

Now that the curve equations for the two mirrors is obtained in their general form, some specific optical layouts for shaping can be achieved by tuning the values of $a$, $b$, and $c$. Here two representative layouts are illustrated in Fig. 2.

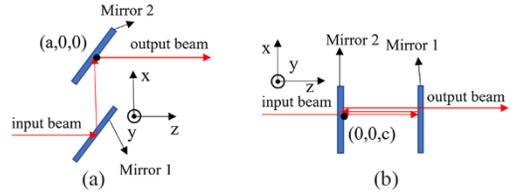

Fig. 2. Two representative layouts for two-mirrors system: (a) 45 degrees of incidence for $b = 0$ and $c = 0$ ; (b) Normal incidence for $a = 0$ and $b = 0$, where the separated rays between two mirrors is just for clarity purposes.

a) 45 degrees of incidence for $b = 0$ and $c = 0$

When $b = 0$ and $c = 0$, the beam is incident at 45 degrees on both free-form mirrors. Under this condition, Eq. 10 and Eq. 13 can be simplified as

$$\begin{cases} \frac{\partial z(r,\theta)}{\partial r} = \frac{-r + R(r) + a\cos\theta}{a} \\ \frac{\partial z(r,\theta)}{\partial \theta} = \frac{-ar\sin\theta}{a} \end{cases} \quad (14)$$

$$\begin{cases} \frac{\partial Z(R,\Theta)}{\partial R} = \frac{-r(R) + R + a\cos\Theta}{a} \\ \frac{\partial Z(R,\Theta)}{\partial \Theta} = \frac{-aR\sin\Theta}{a} \end{cases} \quad (15)$$

The expressions of $z(r,\theta)$ and $Z(R,\Theta)$ are

$$\begin{cases} z(r_0, \theta_0) = -\frac{1}{a} \int_0^{r_0} (r - R(r)) dr + r_0 \cos\theta_0 \\ Z(R_0, \Theta_0) = -\frac{1}{a} \int_0^{R_0} (r(R) - R) dR + R_0 \cos\Theta_0 \end{cases} \quad (16)$$

From Eq. 16, it is evident that the expressions for $z(r,\theta)$ and $Z(R,\Theta)$ consist of two components: the integral term and the angle term. In conjunction with the beam path diagram, it becomes apparent that the angle term only serves to the deviation from its intended angle. Therefore, the integral term should be viewed as the primary contribution to beam shaping.

The shapes of the two aspheric mirrors are not only influenced by $r$ and $R$, but also by $\theta$ and $\Theta$, resulting in a non-axial symmetry of the two mirrors.

b) Normal incidence of $a = 0$ and $b = 0$

It is evident that adjusting the non-axisymmetric mirror form involves more complexity than adjusting the axisymmetric aspheric mirror form. Furthermore, the axisymmetric mirror is easier to process, resulting in a higher surface accuracy. By setting $a = 0$ and $b = 0$, it can be deduced that the surface equations of the mirrors at this time are axisymmetric, as shown in Eq. 17 and Eq. 18

$$\begin{cases} \dfrac{\partial z}{\partial r} = \dfrac{-r + R(r)}{2c} \\ \dfrac{\partial z}{\partial \theta} = 0 \end{cases} \quad (17)$$

$$\begin{cases} \dfrac{\partial Z}{\partial R} = \dfrac{-r(R) + R}{2c} \\ \dfrac{\partial z}{\partial \theta} = 0 \end{cases} \quad (18)$$

Integrate Eq. 17 and Eq. 18, we can obtain

$$\begin{cases} z(r_0) = \int_{(0,0)}^{(r_0,0)} \dfrac{dz}{dr} dr = -\dfrac{1}{2c}\int_0^{r_0}(r - R(r))dr \\ Z(R_0) = \int_{(0,0)}^{(R_0,0)} \dfrac{dZ}{dR} dR = -\dfrac{1}{2c}\int_0^{R_0}(r(R) - R)dR \end{cases} \quad (19)$$

When $a = 0$ and $b = 0$, the beam path is normal incident to shaping system and free-form mirrors can be simplified to AAMs, which are shown in Fig. 2(b). However, since the laser beam is incident perpendicularly on both mirrors, additional optical components are required to separate the output shaped beam from the input beam.

## 5. Simulation and Verification of AAMs-shaping

To realize the separation of the incident unshaped beam and the output shaped beam, the polarization is manipulated when the laser passing through two AAMs, as shown in Fig. 3. Here a polarizing beam splitter (PBS) reflects s-polarized beams while transmitting p-polarized beams, and two quarter-wave plates (QWPs) are employed to convert s-polarized beams into p-polarized beams and vice versa. A combination of a mirror and a QWP has the capability to induce a polarization state transition of the linearly polarized light. This approach ensures the accomplishment of beam shaping by utilizing AAMs, ensuring that the laser beam can vertically incident on two AAMs.

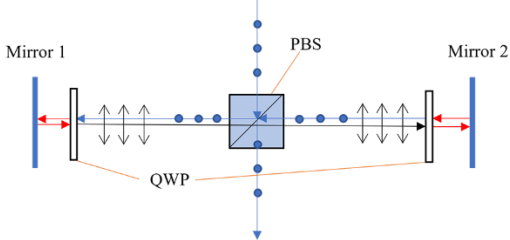

Fig. 3. AAMs-shaping system based on polarization control. The blue (black) arrow represents s-polarized (p-polarized) beam and the red arrow represents circularly polarized beam, respectively.

Restricted by laser-optical processing technology, AAMs exist unavoidable surface roughness, thus leading to intrinsic errors for shaped profiles. In order to accurately model the impact of mirror errors on shaping outcomes, errors with Peak-to-Valley (PV) values of 1nm, 10nm, 100nm, and 1μm are systematically introduced, where the corresponding root-mean-square (RMS) values are 0.3nm, 3nm, 30nm, and 300nm, respectively. Figure 4 illustrates four flat-top shaping results with different PV accuracies of AAMs based on the simulation results. One can see that the flat-top shaping can be achieved with the PV values less than 100nm.

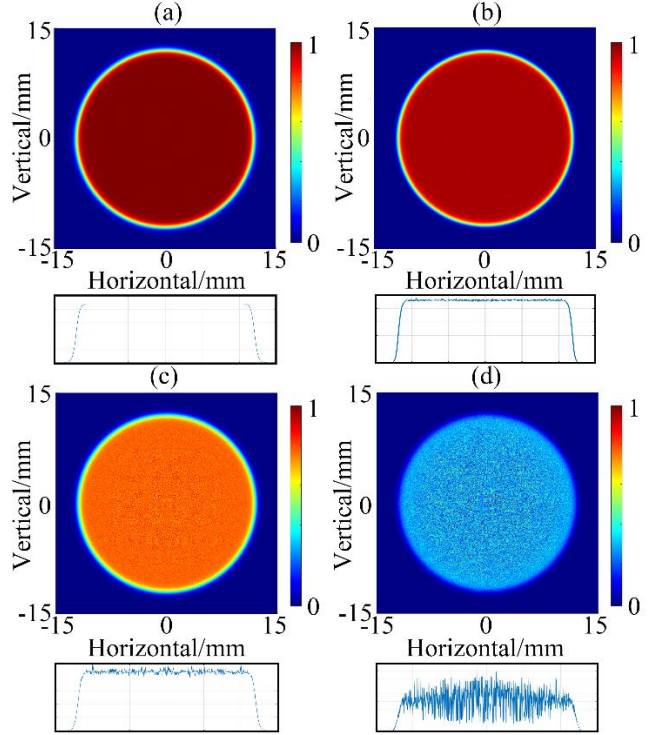

Fig. 4. Impact of varied machining accuracy on beam shaping results:(a) PV Value: 1nm (b) PV Value: 10nm (c) PV Value: 100nm (d) PV Value: 1μm Here q is set to 50 and $R_{FL}$ is fixed to 12mm.

To quantify the disparity between the actual beam and the ideal beam, the root-mean-square error (RMSE) is employed and mathematically expressed as

$$RMSE = \sqrt{\dfrac{1}{N-1}\sum_i^N \left(\dfrac{A_i(r) - B_i(r)}{B_i(r)}\right)^2} \quad (20)$$

Here, $A(r)$ represents the actual intensity distribution, while $B(r)$ represents the ideal intensity distribution. The corresponding RMSE values for PV accuracies of 1nm, 10nm, 100nm, and 1μm are 0.26%, 5.61%, 9.96% and 42.63%, respectively. Thus, without considering the initial error in the beam spot size, to maintain the RMSE value below 10%, it is imperative to ensure that the mirror processing error remains below the PV value of 100nm. Indeed, the simplified machining process and enhanced controllability of machining accuracy inherent to AAMs render the attainment of such machining precision notably more achievable.

Apart from mirror errors, the size errors of the input beam also play a significant role in influencing the shaping outcomes of the beam. Seen from Fig.4(a) and Fig.4(b), a perfect shaped flat-top beam with FWHM beam radius of 12mm can be transformed from an initial Gaussian beam with waist radius of 6mm. Figure 4(c) and Figure 4(d) show two different waist radius of the input beam, respectively. In the simulation, the accuracy of both AAMs is upheld to the 10nm level, with no additional mirror errors incorporated.

The RMSE for Fig. 4(b), 4(c), and 4(d) can be computed as 3.46%, 18.97% and 14.67%, respectively. Therefore, under the assumption of ignoring the processing error of AAMs, if the spot size error is controlled not less than 8%, the RMSE of the light intensity distribution can be kept below 20%. This rigorous control over the spot size error is indispensable for achieving the desired accuracy during beam shaping. Equation 8 reveals that the relationship between $R/R_{FL}$ and $r^2/w^2$ is significant, thereby allowing for the extrapolation of the relative data of 8% and 20% to all parameters.

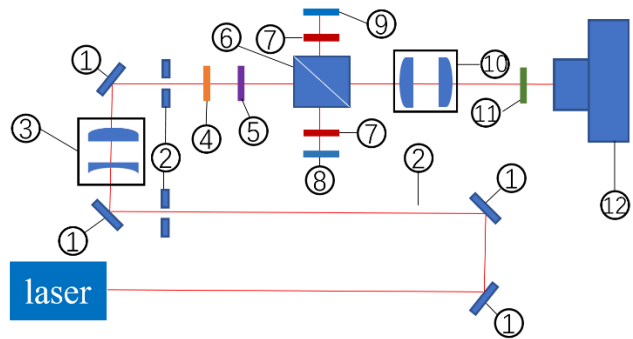

Fig. 6. The AAMs-shaping experimental setup. 1- protected silver mirrors, 2-iris, 3-lenses used for beam expansion, 4-HWP, 5-polarizer, 6-PBS, 7-QWP, 8-the first AAM, 9-the second AAM, 10- lenses used for image transmission, 11- optical filter, 12-CCD camera.

Figure 7(a) shows the intensity distribution of the input laser beam after expanding beam waist radius from 1.21mm to 6.05mm, while Fig. 7(b) shows the intensity distribution of the laser beam after shaping, located at 3cm away from the PBS.

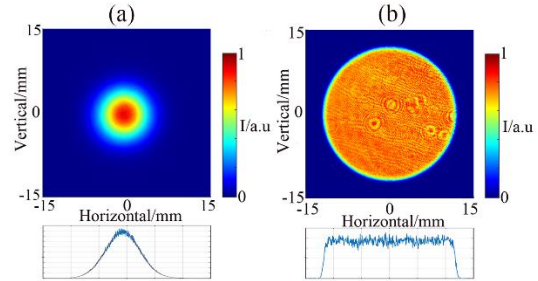

Fig. 7. Experiment results of beam shaping: (a) The intensity distribution of the input beam. (b) The intensity distribution of the output beam

Figure 7 shows the RMSEs for the input and output beams are 3.02% and 12.31%, respectively, in good agreement with the simulation result with the PV value of 100nm (the rms value of 30nm). Obviously, the machining error of the two aspheric mirrors and the nonideal Gaussian input beam contribute to the increase of RMSE. Additionally, the energy efficiency of the beam shaping system is calculated by comparing the number counted by CCD before and after beam shaping with the same filters, resulting in an energy efficiency of ~74%. Multiple factors limit the lossless transmission during the flat-top beam shaping, including the reflection loss associated with AAMs and the transmission loss of polarization manipulation optics.

To study the intensity evolution of the shaped flat-top distribution along the beam propagation, a numerical calculation based on Fresnel diffraction equation is conducted. Figure 8 show the measured intensity evolutions of the shaped beams for the propagation distances of 2.5m and 4m in air respectively, entirely consistent with the simulation results.

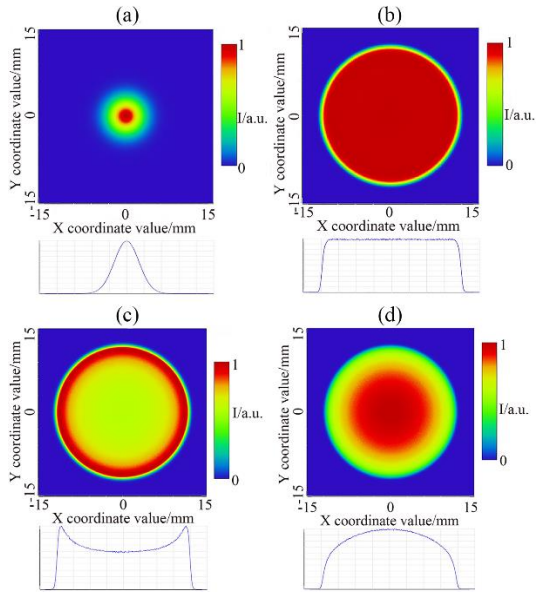

Fig. 4. Simulation results of flat-top beam shaping from various input beam sizes of Gaussian beam: (a) The intensity distribution of the input beam; (b-d) The desired intensity distribution of the output beams for various input beams sizes of (b)$w = 6mm$/(c)$w = 6.5mm$/(d)$w = 5.5mm$.

6. Experimental Demonstration

To experimentally investigate the shaping of flat-top beam from Gaussian beam, two axisymmetric aspheric mirrors were machined by a single point diamond turning method with machining accuracy of RMS 1/10λ at 532nm.

The experimental setup for beam shaping and diagnostic is shown in Fig. 6, where a He-Ne laser is utilized. An image transmission is employed to match the entire shaped beam with CCD chip size.

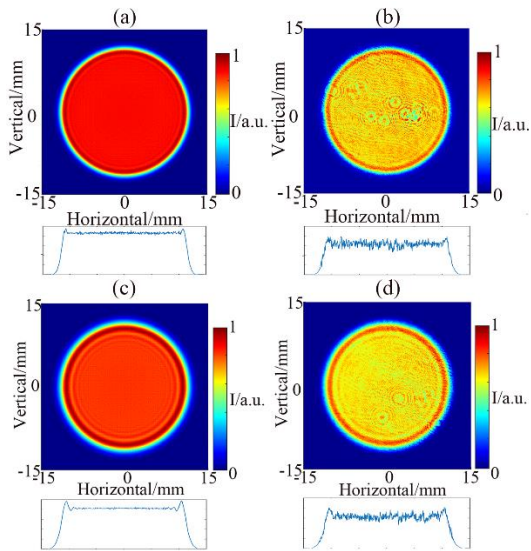

Fig. 8. The evolution of flat-top beam intensity distribution after propagating 2.5m (a, b) and 4m (c, d) in air, respectively, with the comparison between the simulated profiles (a, c) and experimental profiles (b, d).

Since this shaping rely heavily on the polarization manipulation, stringent requirements on the performance of polarization optics are necessary, such as a high extinction ratio and a precise working angle. Failure to achieve these may result in portions of the beam oscillating continuously between the two AAMs, thereby causing noticeable diffraction stripes in the shaped beam. An additional isolator can inhibit the two-beam diffraction by allow forward beam to pass through and isolate reverse beam, as shown in Fig. 9.

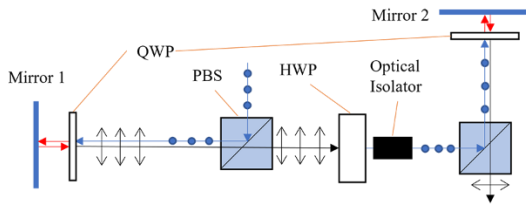

Fig. 9. An improved shaping system by inserted an isolator

7. Conclusion

In this paper, a beam shaping method based on AAMs and polarization manipulation has been experimental demonstrated with efficiency of more than 70%, alleviating the stringent demands of the mirror processing and adjustment in universe free-form mirrors. The RMSE value to evaluate the effect of flat-top distribution is calculated to 12.3%, coincide with the simulated result for the influence of the AAMs surface accuracy. According the simulated verification, the flat-top shaping can be easily achieved with the RMSE value<10% when the machining errors with Peak-to-Valley (PV) values<100nm. Both simulation and experiment results show that the beam size of the input Gaussian beam has a great influence on the expected beam shaping, so that the beam size should be as close as possible to the specific value. This method based on axisymmetric aspheric mirrors paves the way to convenient conversion between various beam profiles.

Acknowledgement
The authors would like to thank Dr. Xianglong Mao for invaluable assistance in AAMs fabrication, Professor Chih-Hao Pai for providing precious advice on beam shaping, Dr. Te-Sheng Hung for useful discussion during experimental demonstration.
This work was support by the National Natural Science Foundation of China (Grant No. 11991071, No. 11991073, No. 12375241 and No. 123051520), the National Science and Technology Major Project (Grant No. 2019-VII-0019-0161), Discipline Construction Foundation of "Double World-class Project", the Strategic Priority Research Program of the Chinese Academy of Sciences (Grant No. XDB0530000). The simulation work is supported by Center of High performance computing, Tsinghua University.